\documentclass[11pt]{article}

\usepackage{hyperref}
\usepackage[a4paper,text={6in,9in},centering]{geometry} 
\usepackage{amsmath}
\usepackage{amssymb}
\usepackage{amsbsy}
\usepackage{mathrsfs}
\usepackage{graphicx}
\usepackage{subfigure}
\usepackage{epic,eepic,xcolor} 	
\usepackage{cite}
\makeatletter

\@addtoreset{equation}{section}
\makeatother

\title{
	\vskip -3em
	\begin{flushright}
	\small
	HRI-P-12-01-002
	\end{flushright}
	\vskip 5em
	\textbf{\large Path Integral Junctions}
}
\author{
Satoshi Ohya
\\[1ex]
\textit{\small Harish-Chandra Research Institute}\\
\textit{\small Chhatnag Road, Jhusi, Allahabad 211 019, India}\\[1ex]
\texttt{\small E-mail:\href{mailto:ohya@hri.res.in}{ohya@hri.res.in}}
}
\date{\small (Dated: \today)}

\begin{document}
\maketitle
\thispagestyle{empty}
\begin{abstract}
We propose path integral description for quantum mechanical systems on compact graphs consisting of $N$ segments of the same length.
Provided the bulk Hamiltonian is segment-independent, scale-invariant boundary conditions given by self-adjoint extension of a Hamiltonian operator turn out to be in one-to-one correspondence with $N \times N$ matrix-valued weight factors on the path integral side.
We show that these weight factors are given by $N$-dimensional unitary representations of the infinite dihedral group.
\end{abstract}

\newpage
\section{Introduction} \label{sec:1}

\begin{figure}[b]
\begin{center}
\begin{tabular}{ccc}
\begin{minipage}[b]{6cm}
\centerline{\includegraphics{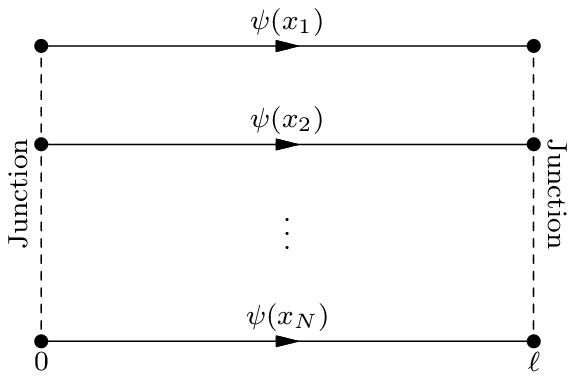}}
\end{minipage} &
\raisebox{1.7cm}{$\xrightarrow[]{\text{folding}}$} &
\begin{minipage}[b]{6cm}
\centerline{\raisebox{1.5cm}{\includegraphics{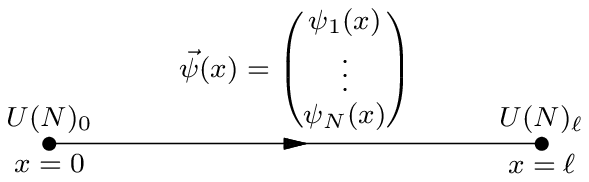}}}
\end{minipage}
\end{tabular}
\caption{Junctions of $N$ segments of the same length $\ell$. Dashed lines represent the current flow among different segments. Arrow indicates the direction of coordinates.}
\label{fig:1}
\end{center}
\end{figure}

It has been long known that, in path integral on multiply-connected space $\mathcal{M}$, linear combination coefficients (weight factors) for homotopically distinct sectors are given by unitary representations of the fundamental group $\pi_{1}(\mathcal{M})$.
For standard ``scalar'' quantum mechanics, i.e. the case where wave functions are given by single-component scalar-valued functions, these are one-dimensional unitary representations \cite{Schulman:1968yv,Laidlaw:1970ei,Schulman:1971,Dowker:1972np} and hence characterized by $U(1)$ phase.
As first discussed by Horvathy \textit{et al.} \cite{Horvathy:1988vh}, however, for ``vector'' quantum mechanics, i.e. the case where wave functions become multi-component vector-valued functions, weight factors could be higher-dimensional unitary representations \cite{Horvathy:1988vh} and hence in principle enjoy much more parameters than those of ``scalar'' quantum mechanics case.

\begin{figure}[t]
\begin{center}
\subfigure[Circle]{
\begin{tabular}{ccc}
\begin{minipage}[b]{3.5cm}
\centerline{\includegraphics{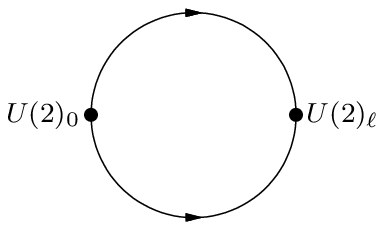}}
\end{minipage} &
\raisebox{1cm}{$\xrightarrow[]{\text{folding}}$} &
\begin{minipage}[b]{3.5cm}
\centerline{\hspace{-.2cm}\raisebox{1cm}{\includegraphics{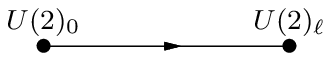}}}
\end{minipage}
\end{tabular} \label{fig:2a}} \\[1em]
\subfigure[Star]{
\begin{tabular}{ccc}
\begin{minipage}[b]{3.5cm}
\centerline{\includegraphics{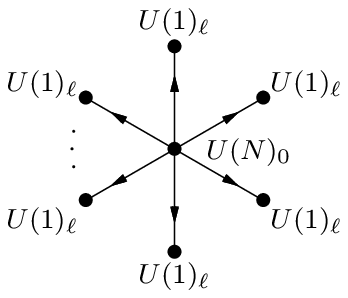}}
\end{minipage} & 
\raisebox{1.35cm}{$\xrightarrow[]{\text{folding}}$} &
\begin{minipage}[b]{3.5cm}
\centerline{\hspace{.1cm}\raisebox{1.35cm}{\includegraphics{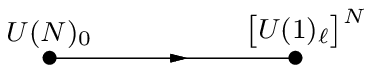}}}
\end{minipage}
\end{tabular} \label{fig:2b}} \\[1em]
\subfigure[Flower]{
\begin{tabular}{ccc}
\begin{minipage}[b]{3.5cm}
\centerline{\includegraphics{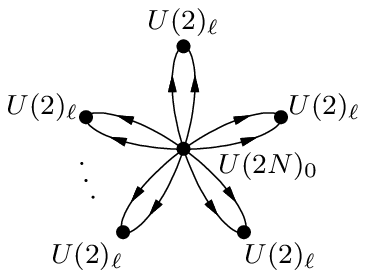}}
\end{minipage} &
\raisebox{1.1cm}{$\xrightarrow[]{\text{folding}}$} &
\begin{minipage}[b]{3.5cm}
\centerline{\raisebox{1.1cm}{\includegraphics{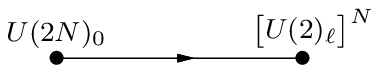}}}
\end{minipage}
\end{tabular} \label{fig:2c}}
\caption{Typical segment junctions and its foldings. (a) the case of $N = 2$, (b) the case of the reduction $U(N)_{\ell} \to [U(1)_{\ell}]^{N}$, and (c) the case of the reduction $U(2N)_{\ell} \to [U(2)_{\ell}]^{N}$.}
\label{fig:2}
\end{center}
\end{figure}

In the present paper we attempt to formulate path integral description of boundary conditions provided by Kirchhoff's law of probability current \cite{Kostrykin:1998} (self-adjoint extension of Hamiltonian operator) on quantum graphs, by continuing our previous investigation on the path integral on star graphs \cite{Ohya:2011qu}.
The key to understanding the path integral description of self-adjoint extension parameters is higher-dimensional unitary representations.
Originally, Horvathy \textit{et al.} focused on internal symmetries as the origin of vector-valued wave functions \cite{Horvathy:1988vh}.
However, the origin of vector-valuedness is not necessarily internal symmetries.
An alternative realization of vector-valued wave functions is folding trick \cite{Oshikawa:1996,Bachas:2001vj}, which maps a defect system to a boundary system.
In this paper we will work on a compact graph that consists of $N$ segments of the same length $\ell$, $(0, \ell) \ni x_{i}$ ($i=1,\cdots,N$), each of which is connected at $x_{i} = 0$ and $\ell$, respectively; see figure \ref{fig:1}.
Such graph can be folded into a system on a single segment with $N$-component vector-valued wave functions, with $U(N)$ family of boundary conditions at each boundary.
An important point to note is that this graph contains a large number of topologically distinct compact graphs as its ``subgroups''; namely, a reduction $U(N) \to U(M_{1}) \times \cdots \times U(M_{n})$ ($M_{1} + \cdots + M_{n} = N$) leads to a number of compact graphs that have different topologies, which is nothing but the topology change discussed in \cite{Balachandran:1995jm,Asorey:2004kk}.
Typical examples of such compact graphs are depicted in figure \ref{fig:2}.

A puzzle may arise from homotopy structures in folded theories, however.
Typically, folded systems boil down to systems on half-line or segment, which are simply-connected spaces such that its fundamental groups are trivial.
Nonetheless, path integral on, say, a segment is known to be given by a sum of partial amplitudes corresponding to an infinite number of classical bouncing paths (for textbook exposition see e.g. chapter 6 of \cite{Kleinert:2009}).
Actually, this is an evidence that such system also intrinsically possesses its own group structure in the path space.

The purpose of the paper is twofold.
Firstly, we would like to clarify the group structure of the space of all possible paths on a segment.
Although the path integral on a segment has been studied over the years \cite{Janke:1979,Goodman:1981,Carreau:1990wh,Nevels:1993}, to the best of our knowledge, little is known about its group structure.
We show that the path space on a segment is furnished with the structure of the infinite dihedral group $D_{\infty} = \mathbb{Z} \rtimes \mathbb{Z}_{2}$.
Secondly, we discuss that path integral on a segment junction is given by a linear combination of partial amplitudes, each of which corresponds to a distinct $D_{\infty}$ sector.
We show that, for the case of scale-invariant subfamily of boundary conditions, the weight factor for each $D_{\infty}$ sector falls into an $N$-dimensional unitary representation of $D_{\infty}$, which is the main result of the present paper.

The organization of the paper is as follows:
We will begin in section \ref{sec:2} by studying a system for a free spinless particle on segment junctions in the folding picture, which is exactly solvable and exemplifies enough the path structure of the system.
This non-interacting theory provides the simplest yet concrete illustration for the group theoretical structure of the weight factors.
In section \ref{sec:3} we generalize to interacting theories by introducing segment-independent bulk interactions.
We show that the weight factors in the path integral are generally given by $N$-dimensional unitary representations of the infinite dihedral group $D_{\infty}$.
Section \ref{sec:4} is devoted to conclusions.
Computational details are relegated to appendix \ref{appendix:A}.

Throughout the paper we will work in the units $\hbar = 2m = 1$.

\section{Scale-invariant boundary conditions} \label{sec:2}
Let us first start with quantum mechanics for a free particle on junctions of $N$ segments of the same length $\ell$.
By folding trick \cite{Oshikawa:1996,Bachas:2001vj}, such system is mapped to an alternative equivalent system on a single segment $(0, \ell)$, whose Hilbert space is given by a tensor product $\mathcal{H} = L^{2}(0, \ell) \otimes \mathbb{C}^{N}$.
There, the wave function becomes an $N$-component vector-valued function $\Vec{\psi}(x) = (\psi_{1}(x), \cdots, \psi_{N}(x))^{T} \in \mathcal{H}$; see figure \ref{fig:1}.
In this system self-adjoint extension of free Hamiltonian $\mathbb{H}_{\text{free}} = -\frac{d^{2}}{dx^{2}} \otimes \mathbb{I}_{N}$ leads to the following $U(N)_{0} \times U(N)_{\ell}$ family of boundary conditions \cite{Kuchment:2008}:
\begin{subequations}
\begin{align}
(\mathbb{I}_{N} - \mathbb{U}_{0})\Vec{\psi}(0)
- iL(\mathbb{I}_{N} + \mathbb{U}_{0})\Vec{\psi}^{\prime}(0)
&= 	\Vec{0}, \quad \mathbb{U}_{0} \in U(N)_{0}, \label{eq:2.1a}\\
(\mathbb{I}_{N} - \mathbb{U}_{\ell})\Vec{\psi}(\ell)
+ iL(\mathbb{I}_{N} + \mathbb{U}_{\ell})\Vec{\psi}^{\prime}(\ell)
&= 	\Vec{0}, \quad \mathbb{U}_{\ell} \in U(N)_{\ell}, \label{eq:2.1b}
\end{align}
\end{subequations}
where $\mathbb{I}_{N}$ is an $N \times N$ identity matrix, $L$ is an arbitrary length scale and $\Vec{\psi}^{\prime}(x) = \frac{d\Vec{\psi}}{dx}(x)$.
The sign difference $\pm iL$ is just for later convenience.
We note that figure \ref{fig:2a}, \ref{fig:2b}, and \ref{fig:2c} correspond, respectively, to the cases of $N = 2$, $U(N)_{\ell} \to [U(1)_{\ell}]^{N}$, and $U(N)_{\ell} \to [U(2)_{\ell}]^{N/2}$ ($N$: even integer).

We now focus on the case of scale-invariant boundary conditions, which are specified by hermitian unitary matrices \cite{Cheon:2010,Ohya:2011qu} that satisfy $\mathbb{U}_{0}^{2} = \mathbb{U}_{\ell}^{2} = \mathbb{I}_{N}$.
In that case the boundary conditions \eqref{eq:2.1a} \eqref{eq:2.1b} are simply reduced to the following conditions:
\begin{subequations}
\begin{align}
&(\mathbb{I}_{N} - \mathbb{U}_{0})\Vec{\psi}(0)
= 	(\mathbb{I}_{N} + \mathbb{U}_{0})\Vec{\psi}^{\prime}(0)
= 	\Vec{0}, \label{eq:2.2a}\\
&(\mathbb{I}_{N} - \mathbb{U}_{\ell})\Vec{\psi}(\ell)
= 	(\mathbb{I}_{N} + \mathbb{U}_{\ell})\vec{\psi}^{\prime}(\ell)
= 	\Vec{0}, \label{eq:2.2b}
\end{align}
\end{subequations}
which follow from multiplying equations \eqref{eq:2.1a} and \eqref{eq:2.1b} by $\mathbb{I}_{N} \pm \mathbb{U}_{0}$ and $\mathbb{I}_{N} \pm \mathbb{U}_{\ell}$, respectively.
Note that the boundary conditions \eqref{eq:2.2a} and \eqref{eq:2.2b} do not involve the scale parameter $L$, as it should.
Now it is easy to solve the Schr\"odinger equation $\mathbb{H}_{\text{free}}\Vec{\psi}(x) = E\Vec{\psi}(x)$ and find the complete orthonormal set of energy eigenfunctions with respect to \eqref{eq:2.2a} and \eqref{eq:2.2b}.
By expanding the kernel in terms of the energy eigenfunctions and then using the argument principle, which enables us to switch from the summation over energy spectrum to the summation over (classical) trajectories, the total Feynman kernel $\mathbb{K}(x, y; T) = \langle x|\mathrm{e}^{-i\mathbb{H}_{\text{free}}T}|y\rangle$ is cast into the following $N \times N$ matrix-valued form (see appendix \ref{appendix:A}):
\begin{align}
\mathbb{K}(x, y; T)
&= 	\sum_{n\in\mathbb{Z}}
	(\mathbb{U}_{0}\mathbb{U}_{\ell})^{n}
	K_{\text{free}}(x, y-2n\ell; T) \nonumber\\
& 	+ \sum_{n\in\mathbb{Z}}
	(\mathbb{U}_{0}\mathbb{U}_{\ell})^{n}\mathbb{U}_{0}
	K_{\text{free}}(x, -y-2n\ell; T), \quad 0 < x, y < \ell, \label{eq:2.3}
\end{align}
whose $ij$-component gives the transition amplitude for a particle propagating from the position $y$ on the $j$th segment to the position $x$ on the $i$th segment.
$K_{\text{free}}(x, y; T)$ is the scalar Feynman kernel for a single free particle on a whole line $\mathbb{R}$ and given by $K_{\text{free}}(x, y; T) = \frac{1}{\sqrt{4\pi iT}}\exp\bigl[i\frac{T}{4}\bigl(\frac{x - y}{T}\bigr)^{2}\bigr]$.
Notice that the scalar kernel is invariant under the action of both reflection and continuous translation, $K_{\text{free}}(-x, -y; T) = K_{\text{free}}(x, y; T)$ and $K_{\text{free}}(x + a, y + a; T) = K_{\text{free}}(x, y; T)$ ($a$: arbitrary constant).
We also note that the total Feynman kernel \eqref{eq:2.3} satisfies the following conditions:
\begin{subequations}
\begin{alignat}{3}
&\text{(Initial condition)}&\quad
&\mathbb{K}(x, y; 0) = \delta(x - y)\mathbb{I}_{N};& \label{eq:2.4a}\\
&\text{(Composition rule)}&\quad
&\int_{0}^{\ell}\!\!\!dz\,\mathbb{K}(x, z; T_{1})\mathbb{K}(z, y; T_{2})
= \mathbb{K}(x, y; T_{1} + T_{2});& \label{eq:2.4b}\\
&\text{(Unitarity)}&\quad
&\mathbb{K}^{\dagger}(x, y; T) = \mathbb{K}(y, x; -T);& \label{eq:2.4c}\\[1.5ex]
&\text{(Boundary conditions)}&\quad
&(\mathbb{I}_{N} - \mathbb{U}_{0})\mathbb{K}(0, y; T) = (\mathbb{I}_{N} + \mathbb{U}_{0})(\partial_{x}\mathbb{K})(0, y; T) = 0,& \label{eq:2.4d}\\[1.5ex]
& &
&(\mathbb{I}_{N} - \mathbb{U}_{\ell})\mathbb{K}(\ell, y; T) = (\mathbb{I}_{N} + \mathbb{U}_{\ell})(\partial_{x}\mathbb{K})(\ell, y; T) = 0.& \label{eq:2.4e}
\end{alignat}
\end{subequations}

\begin{figure}[t]
\begin{center}
\subfigure[Classical trajectory (weight factor: $\mathbb{U}_{0}\mathbb{U}_{\ell}$)]
{\includegraphics{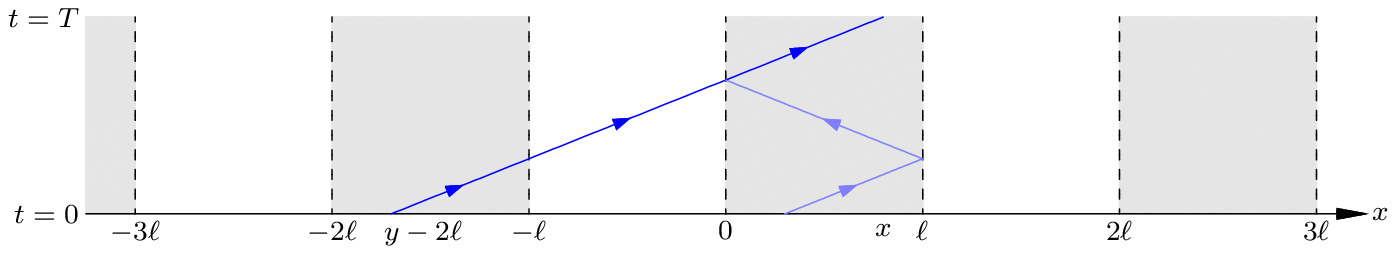} \label{fig:3a}}
\subfigure[Classical trajectory (weight factor: $\mathbb{U}_{0}\mathbb{U}_{\ell}\mathbb{U}_{0}$)]
{\includegraphics{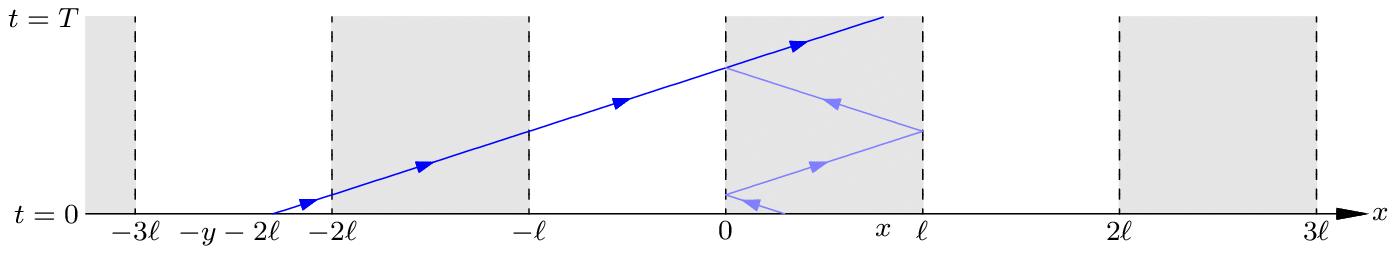} \label{fig:3b}}
\subfigure[Quantum fluctuation of (a) (weight factor: $\mathbb{U}_{\ell}\mathbb{U}_{\ell}\mathbb{U}_{0}\mathbb{U}_{\ell} = \mathbb{U}_{0}\mathbb{U}_{\ell}$)]
{\includegraphics{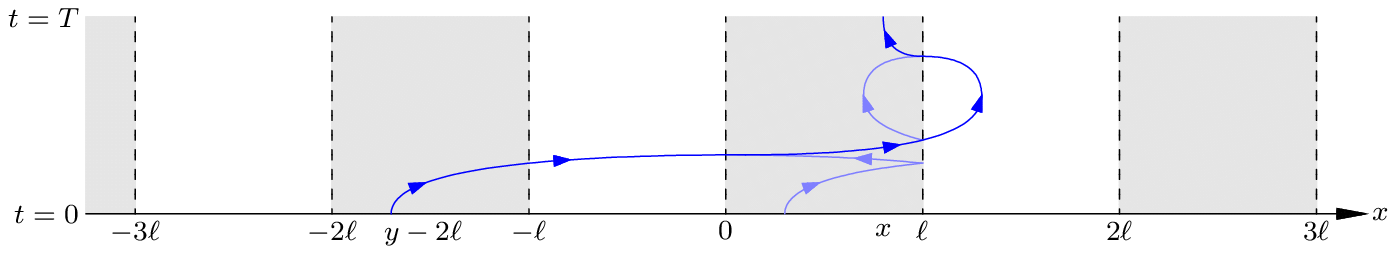} \label{fig:3c}}
\subfigure[Quantum fluctuation of (b) (weight factor: $\mathbb{U}_{0}\mathbb{U}_{\ell}\mathbb{U}_{\ell}\mathbb{U}_{\ell}\mathbb{U}_{0} = \mathbb{U}_{0}\mathbb{U}_{\ell}\mathbb{U}_{0}$)]
{\includegraphics{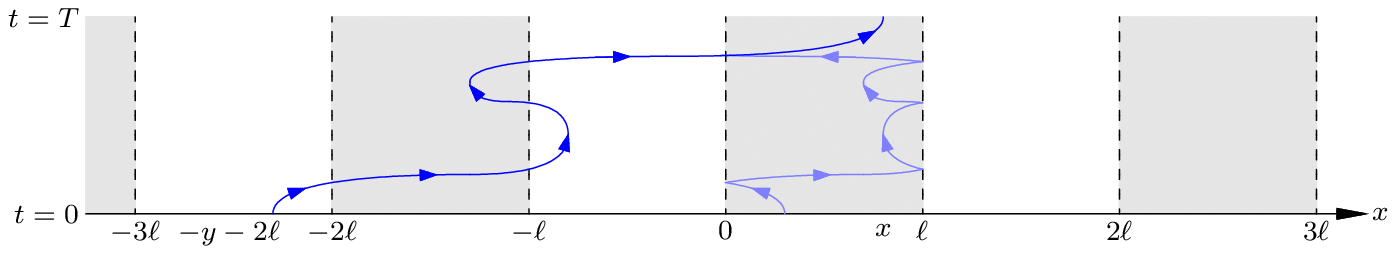} \label{fig:3d}}
\caption{Typical classical trajectories and its quantum fluctuations.}
\label{fig:3}
\end{center}
\end{figure}

We now turn to the physical interpretation of the total Feynman kernel \eqref{eq:2.3}.
The exponent $\frac{T}{4}\bigl[\frac{x - (\pm y - 2n\ell)}{T}\bigr]^{2}$ is nothing but the action integral $S[x_{\text{cl}}] = \int_{0}^{T}\!\!dt\,\frac{1}{4}\Dot{x}_{\text{cl}}^{2}(t)$ for a free particle on $\mathbb{R}$ traveling along the classical trajectory $x_{\text{cl}}(t) = x_{i} + (\tfrac{x_{f} - x_{i}}{T})t$ with initial and final positions $x_{i} = \pm y - 2n\ell$ and $x_{f} = x$.
Thanks to the translation- and reflection-invariance of the scalar kernel, these straight line trajectories on $\mathbb{R}$ can be interpreted as bouncing paths on the segment $(0, \ell)$; see figure \ref{fig:3}.
These bouncing paths are weighted by the scale-invariant S-matrix $\mathbb{U}_{0}$ ($\mathbb{U}_{\ell}$) every time when a particle bounces off the boundary $x = 0$ ($x = \ell$).
($\mathbb{U}_{0}$ and $\mathbb{U}_{\ell}$ play the roles of boundary S-matrices; see appendix \ref{appendix:A}.)
For example, for a classical straight line trajectory with initial position $y - 2n\ell$ ($n>0$), which corresponds to the bouncing path that hits the boundaries $2n$ times in alternating order, the weight factor is given by an alternating product of hermitian unitary S-matrices $\mathbb{U}_{0}\mathbb{U}_{\ell} \cdots \mathbb{U}_{0}\mathbb{U}_{\ell} = (\mathbb{U}_{0}\mathbb{U}_{\ell})^{n}$; see figure \ref{fig:3a} for the case of $n=1$.
Likewise, for a classical trajectory with initial position $- y - 2n\ell$ ($n > 0$), which corresponds to the bouncing path that hits the boundaries $2n + 1$ times in alternating order, the weight factor is given by $\mathbb{U}_{0}\mathbb{U}_{\ell} \cdots \mathbb{U}_{0}\mathbb{U}_{\ell}\mathbb{U}_{0} = (\mathbb{U}_{0}\mathbb{U}_{\ell})^{n}\mathbb{U}_{0}$; see figure \ref{fig:3b} for the case of $n=1$.
The summation over all integer $n$ thus ensures the summation over all possible bouncing paths on the segment.

\section{\texorpdfstring{Weight factor and $N$-dimensional unitary representation of $D_{\infty} \cong \mathbb{Z}_{2} \ast \mathbb{Z}_{2}$}
{Weight factor and N-dimensional unitary representation of infinite dihedral group}} \label{sec:3}
So far we have seen that the hermitian unitary matrices $\mathbb{U}_{0}$ and $\mathbb{U}_{\ell}$ which specify the scale-invariant boundary conditions on the operator formalism side have appeared as weight factors on the path integral side.
In this section we first clarify the group structure of the path space on a segment and then show that, without invoking the boundary conditions, $\mathbb{U}_{0}$ and $\mathbb{U}_{\ell}$ are introduced via $N$-dimensional unitary representations of the infinite dihedral group.
This algebraic consideration leads to the generalization to systems in the presence of bulk interactions.

To begin with, let us first recall basics of the infinite dihedral group $D_{\infty} = \mathbb{Z} \rtimes \mathbb{Z}_{2}$, which is generated by two operations, a discrete translation $\mathcal{T} \in \mathbb{Z}$ and a reflection $\mathcal{R} \in \mathbb{Z}_{2}$, whose actions on $\mathbb{R}$ can be defined as $\mathcal{T}: y \mapsto y - 2\ell$ and $\mathcal{R}: y \mapsto -y$.
All elements of $D_{\infty}$ are then of the forms $\mathcal{T}^{n}$ and $\mathcal{T}^{n}\mathcal{R} (= \mathcal{R}\mathcal{T}^{-n})$ with integer $n$, which act on $\mathbb{R}$ as $\mathcal{T}^{n}: y \mapsto y-2n\ell$ and $\mathcal{T}^{n}\mathcal{R}: y \mapsto -y-2n\ell$, and obey the following multiplication rules:
\begin{subequations}
\begin{alignat}{3}
&\mathcal{T}^{n} \circ \mathcal{T}^{m} = \mathcal{T}^{n+m},&\quad
&\mathcal{T}^{n} \circ \mathcal{T}^{m}\mathcal{R} = \mathcal{T}^{n+m}\mathcal{R},& \label{eq:3.1a}\\
&\mathcal{T}^{n}\mathcal{R} \circ \mathcal{T}^{m} = \mathcal{T}^{n-m}\mathcal{R},&\quad
&\mathcal{T}^{n}\mathcal{R} \circ \mathcal{T}^{m}\mathcal{R} = \mathcal{T}^{n-m},& \label{eq:3.1b}
\end{alignat}
\end{subequations}
where circle ($\circ$) stands for abstract group multiplication.
Note that identity element is $\mathcal{T}^{0}$, and inverse elements are $(\mathcal{T}^{n})^{-1} = \mathcal{T}^{-n}$ and $(\mathcal{T}^{n}\mathcal{R})^{-1} = \mathcal{T}^{n}\mathcal{R}$.
Note also that $\mathcal{T}^{n}$ ($n \neq 0$) and $\mathcal{T}^{m}\mathcal{R}$ are non-commutative, $\mathcal{T}^{n} \circ \mathcal{T}^{m}\mathcal{R} \neq \mathcal{T}^{m}\mathcal{R} \circ \mathcal{T}^{n}$; that is, $D_{\infty}$ is a non-Abelian discrete group.

Let us next clarify the group structure of the space of all possible paths on a segment.
As we have seen in section \ref{sec:2}, the total Feynman kernel on a segment is given by a linear combination of partial amplitudes.
Each partial amplitude is obtained by integration over one particular path class, which is a set of paths on $\mathbb{R}$ that cross each wall between the initial and final positions odd number of times.
Such classification of paths is furnished with the infinite dihedral group structure.
Indeed, as a natural generalization of the classical free particle trajectories, paths shall acquire the S-matrix $\mathbb{U}_{0}$ ($\mathbb{U}_{\ell}$) every time when a particle crosses the walls at $x = -2n\ell$ ($x = -(2n+1)\ell$); see figure \ref{fig:3c} and \ref{fig:3d}.
Any paths shall therefore be labeled by certain product of S-matrices.
Then, thanks to the hermitian unitarity of the S-matrices, $\mathbb{U}_{0}^{2} = \mathbb{I}_{N}$ and $\mathbb{U}_{\ell}^{2} = \mathbb{I}_{N}$, any weight factor falls into an alternating product of $\mathbb{U}_{0}$ and $\mathbb{U}_{\ell}$.
Thus, without any loss of generality, we may denote these path classes as $[(\mathbb{U}_{0}\mathbb{U}_{\ell})^{n}]$ and $[(\mathbb{U}_{0}\mathbb{U}_{\ell})^{n}\mathbb{U}_{0}]$.
Composition of these path classes are given by, for example, $[\mathbb{U}_{0}\mathbb{U}_{\ell}\mathbb{U}_{0}] \ast [\mathbb{U}_{0}\mathbb{U}_{\ell}] = [\mathbb{U}_{0}\mathbb{U}_{\ell}\mathbb{U}_{0}\mathbb{U}_{0}\mathbb{U}_{\ell}] = [\mathbb{U}_{0}]$ and $[\mathbb{U}_{0}\mathbb{U}_{\ell}] \ast [\mathbb{U}_{0}\mathbb{U}_{\ell}\mathbb{U}_{0}] = [\mathbb{U}_{0}\mathbb{U}_{\ell}\mathbb{U}_{0}\mathbb{U}_{\ell}\mathbb{U}_{0}]$.
By making use of the fact that $D_{\infty}$ is isomorphic to the free product $\mathbb{Z}_{2} \ast \mathbb{Z}_{2}$, which we will revisit shortly, we can easily check that these path classes $\{[(\mathbb{U}_{0}\mathbb{U}_{\ell})^{n}], [(\mathbb{U}_{0}\mathbb{U}_{\ell})^{n}\mathbb{U}_{0}]\}$ satisfy the $D_{\infty} \cong \mathbb{Z}_{2} \ast \mathbb{Z}_{2}$ multiplication rules \eqref{eq:3.1a} \eqref{eq:3.1b} under the composition of paths.
A crucial point to note is that these path classes are simply classified by its initial positions $\mathcal{T}^{n}y = y-2n\ell$ and $\mathcal{T}^{n}\mathcal{R}y = -y-2n\ell$.

Now we are ready to discuss the weight factor in path integral on segment junctions.
As the simplest generalization of free particle case discussed in the previous section, we will concentrate on a system described by the segment-independent bulk Hamiltonian that takes the form $\mathbb{H}_{\text{bulk}} = H_{\text{bulk}} \otimes \mathbb{I}_{N}$.
In this case a kernel that acts on $\mathcal{H} = L^{2}(0, \ell) \otimes \mathbb{C}^{N}$ should be factorized as $(\text{scalar kernel}) \times (\text{$N \times N$ constant matrix})$.
As discussed in the previous paragraph, the total Feynman kernel must be given by a linear combination of partial amplitudes, each of which is weighted by an $N \times N$ matrix-valued linear combination coefficient.
Putting these considerations together, the total Feynman kernel must be of the form
\begin{align}
\mathbb{K}(x, y; T)
&= 	\sum_{n\in\mathbb{Z}}
	\bigl[
	\mathbb{W}(\mathcal{T}^{n})K_{\mathbb{R}}(x, \mathcal{T}^{n}y; T)
	+ \mathbb{W}(\mathcal{T}^{n}\mathcal{R})K_{\mathbb{R}}(x, \mathcal{T}^{n}\mathcal{R}y; T)
	\bigr], \quad 0 < x,y < \ell, \label{eq:3.2}
\end{align}
where $\{\mathbb{W}(\mathcal{T}^{n}), \mathbb{W}(\mathcal{T}^{n}\mathcal{R})\}$ are $N \times N$ constant matrices.
$K_{\mathbb{R}}(x, y; T)$ is the scalar kernel on $\mathbb{R}$ and assumed to satisfy the following conditions:
\begin{subequations}
\begin{align}
&\text{(Initial condition)}&\quad
&K_{\mathbb{R}}(x, y; 0) = \delta(x - y);& \label{eq:3.3a}\\
&\text{(Composition rule)}&\quad
&\int_{-\infty}^{\infty}\!\!\!dz\,K_{\mathbb{R}}(x, z; T_{1})K_{\mathbb{R}}(z, y; T_{2})
= K_{\mathbb{R}}(x, y; T_{1} + T_{2});& \label{eq:3.3b}\\
&\text{(Unitarity)}&\quad
&K_{\mathbb{R}}^{\ast}(x, y; T) = K_{\mathbb{R}}(y, x; -T);& \label{eq:3.3c}\\[1.5ex]
&\text{($D_{\infty}$ symmetry)}&\quad
&K_{\mathbb{R}}(x, y; T) = K_{\mathbb{R}}(\mathcal{G}x, \mathcal{G}y; T), \quad \forall \mathcal{G} \in D_{\infty}.& \label{eq:3.3d}
\end{align}
\end{subequations}
We note that $D_{\infty}$ symmetry will be justified by translation- and reflection-invariant bulk scalar Hamiltonian, $\mathcal{T}H_{\text{bulk}}\mathcal{T}^{-1} = \mathcal{R}H_{\text{bulk}}\mathcal{R}^{-1} = H_{\text{bulk}}$.
Below we will show that, under the assumptions \eqref{eq:3.3a}--\eqref{eq:3.3d}, the ansatz \eqref{eq:3.2} will satisfy the initial condition \eqref{eq:2.4a}, composition rule \eqref{eq:2.4b} and unitarity \eqref{eq:2.4c} if and only if the weight factors $\{\mathbb{W}(\mathcal{T}^{n}), \mathbb{W}(\mathcal{T}^{n}\mathcal{R})\}$ are given by $N$-dimensional unitary representations of $D_{\infty}$.
After proving this statement, we will rederive the scale-invariant boundary conditions by using unitary representations.

The proof is almost parallel to that presented in the previous work \cite{Ohya:2011qu} such that we here discuss only the composition rule.
By substituting the ansatz \eqref{eq:3.2}, the left hand side of \eqref{eq:2.4b} becomes
\begin{align}
&\sum_{n,m \in \mathbb{Z}}
\mathbb{W}(\mathcal{T}^{n})\mathbb{W}(\mathcal{T}^{m})
\int_{0}^{\ell}\!\!dz\,
K_{\mathbb{R}}(x, \mathcal{T}^{n}z; T_{1})K_{\mathbb{R}}(z, \mathcal{T}^{m}y; T_{2}) \nonumber\\
&+ \sum_{n,m \in \mathbb{Z}}
\mathbb{W}(\mathcal{T}^{n}\mathcal{R})\mathbb{W}(\mathcal{T}^{m}\mathcal{R})
\int_{0}^{\ell}\!\!dz\,
K_{\mathbb{R}}(x, \mathcal{T}^{n}\mathcal{R}z; T_{1})K_{\mathbb{R}}(z, \mathcal{T}^{m}\mathcal{R}y; T_{2}) \nonumber\\
&+ \sum_{n,m \in \mathbb{Z}}
\mathbb{W}(\mathcal{T}^{n})\mathbb{W}(\mathcal{T}^{m}\mathcal{R})
\int_{0}^{\ell}\!\!dz\,
K_{\mathbb{R}}(x, \mathcal{T}^{n}z; T_{1})K_{\mathbb{R}}(z, \mathcal{T}^{m}\mathcal{R}y; T_{2}) \nonumber\\
&+ \sum_{n,m \in \mathbb{Z}}
\mathbb{W}(\mathcal{T}^{n}\mathcal{R})\mathbb{W}(\mathcal{T}^{m})
\int_{0}^{\ell}\!\!dz\,
K_{\mathbb{R}}(x, \mathcal{T}^{n}\mathcal{R}z; T_{1})K_{\mathbb{R}}(z, \mathcal{T}^{m}y; T_{2}). \label{eq:3.4}
\end{align}
Noting that the identities $K_{\mathbb{R}}(z, \mathcal{T}^{m}y; T_{2}) = K_{\mathbb{R}}(\mathcal{T}^{n}z, \mathcal{T}^{n+m}y; T_{2})$ and $K_{\mathbb{R}}(z, \mathcal{T}^{m}\mathcal{R}y; T_{2}) = K_{\mathbb{R}}(\mathcal{T}^{n}\mathcal{R}z, \mathcal{T}^{n-m}y; T_{2})$, which follow from $D_{\infty}$ symmetry \eqref{eq:3.3d} and $D_{\infty}$ multiplication rules \eqref{eq:3.1a} \eqref{eq:3.1b}, the first two integrals in \eqref{eq:3.4} become $\int_{0}^{\ell}dz\,K_{\mathbb{R}}(x, \mathcal{T}^{n}z; T_{1})K_{\mathbb{R}}(z, \mathcal{T}^{m}y; T_{2}) = \int_{-2n\ell}^{(-2n+1)\ell}d\xi\,K_{\mathbb{R}}(x, \xi; T_{1})K_{\mathbb{R}}(\xi, \mathcal{T}^{n+m}y; T_{2})$ and $\int_{0}^{\ell}dz\,K_{\mathbb{R}}(x, \mathcal{T}^{n}\mathcal{R}z; T_{1})K_{\mathbb{R}}(z, \mathcal{T}^{m}\mathcal{R}y; T_{2}) = \int_{-(2n+1)\ell}^{-2n\ell}d\xi\,K_{\mathbb{R}}(x, \xi; T_{1})K_{\mathbb{R}}(\xi, \mathcal{T}^{n-m}y; T_{2})$, where we have changed the integration variables as $\xi = \mathcal{T}^{n}z = z - 2n\ell$ and $\xi = \mathcal{T}^{n}\mathcal{R}z = -z - 2n\ell$, respectively.
Similarly, the last two integrals in \eqref{eq:3.4} can be written as $\int_{-2n\ell}^{(-2n+1)\ell}d\xi K_{\mathbb{R}}(x, \xi; T_{1})K_{\mathbb{R}}(\xi, \mathcal{T}^{n+m}\mathcal{R}y; T_{2})$ and $\int_{-(2n+1)\ell}^{-2n\ell}d\xi K_{\mathbb{R}}(x, \xi; T_{1})K_{\mathbb{R}}(\xi, \mathcal{T}^{n-m}\mathcal{R}y; T_{2})$.
Collecting the above pieces the left hand side of \eqref{eq:2.4b} is cast into the following form:
\begin{align}
&	\sum_{n,m \in \mathbb{Z}}
	\left[
	\mathbb{W}(\mathcal{T}^{n})\mathbb{W}(\mathcal{T}^{m})
	\int_{-2n\ell}^{(-2n+1)\ell}\!\!\!\!d\xi
	+ \mathbb{W}(\mathcal{T}^{n}\mathcal{R})\mathbb{W}(\mathcal{T}^{-m}\mathcal{R})
	\int_{-(2n+1)\ell}^{-2n\ell}\!\!\!d\xi
	\right] \nonumber\\
&	\times K_{\mathbb{R}}(x, \xi; T_{1})K_{\mathbb{R}}(\xi, \mathcal{T}^{n+m}y; T_{2}) \nonumber\\
&	+ \sum_{n,m \in \mathbb{Z}}
	\left[
	\mathbb{W}(\mathcal{T}^{n})\mathbb{W}(\mathcal{T}^{m}\mathcal{R})
	\int_{-2n\ell}^{(-2n+1)\ell}\!\!\!\!d\xi
	+ \mathbb{W}(\mathcal{T}^{n}\mathcal{R})\mathbb{W}(\mathcal{T}^{-m})
	\int_{-(2n+1)\ell}^{-2n\ell}\!\!\!d\xi
	\right] \nonumber\\
&	\times K_{\mathbb{R}}(x, \xi; T_{1})K_{\mathbb{R}}(\xi, \mathcal{T}^{n+m}\mathcal{R}y; T_{2}), \label{eq:3.5}
\end{align}
where in the second and fourth terms we have changed the summation variable as $m \to -m$.
Notice that the integral $\int_{-2n\ell}^{(-2n+1)\ell}d\xi$ and $\int_{-(2n+1)\ell}^{-2n\ell}d\xi$ cover only the shaded and unshaded regions of figure \ref{fig:3}, respectively.
Thus, in order to realize the right hand side of \eqref{eq:2.4b} with the assumption \eqref{eq:3.3b}, $\mathbb{W}$ must satisfy the following conditions:
\begin{subequations}
\begin{align}
&\mathbb{W}(\mathcal{T}^{n})\mathbb{W}(\mathcal{T}^{m})
= 	\mathbb{W}(\mathcal{T}^{n}\mathcal{R})\mathbb{W}(\mathcal{T}^{-m}\mathcal{R})
= 	\mathbb{W}(\mathcal{T}^{n+m}), \label{eq:3.6a}\\
&\mathbb{W}(\mathcal{T}^{n})\mathbb{W}(\mathcal{T}^{m}\mathcal{R})
= 	\mathbb{W}(\mathcal{T}^{n}\mathcal{R})\mathbb{W}(\mathcal{T}^{-m})
= 	\mathbb{W}(\mathcal{T}^{n+m}\mathcal{R}), \label{eq:3.6b}
\end{align}
\end{subequations}
which are nothing but $D_{\infty}$ multiplication rules \eqref{eq:3.1a} and \eqref{eq:3.1b}.
Hence $\mathbb{W}$ must be an $N \times N$ matrix representation of $D_{\infty}$.
Similarly, one can obtain the conditions for the weight factors by imposing the initial condition \eqref{eq:2.4a} and the unitarity \eqref{eq:2.4c}.
It is straightforward to show that these two conditions lead to the following constraints for the weight factors:
\begin{align}
\mathbb{W}(\mathcal{T}^{0})
&= 	\mathbb{I}_{N}, \label{eq:3.7}
\end{align}
and
\begin{align}
\mathbb{W}^{\dagger}(\mathcal{T}^{n}) = \mathbb{W}(\mathcal{T}^{-n}), \quad
\mathbb{W}^{\dagger}(\mathcal{T}^{n}\mathcal{R}) = \mathbb{W}(\mathcal{T}^{n}\mathcal{R}), \label{eq:3.8}
\end{align}
from which one deduces that the weight factors must be unitary matrices, $\mathbb{W}^{-1}(\mathcal{T}^{n}) = \mathbb{W}^{\dagger}(\mathcal{T}^{n})$ and $\mathbb{W}^{-1}(\mathcal{T}^{n}\mathcal{R}) = \mathbb{W}^{\dagger}(\mathcal{T}^{n}\mathcal{R})$.
Putting all these results together, we find that $\mathbb{W}$ must be an $N$-dimensional unitary representation of the infinite dihedral group
\begin{align}
\mathbb{W}: D_{\infty} \to U(N). \label{eq:3.9}
\end{align}

Now, the problem is reduced to the problem of how to construct such unitary representations.
To solve this, we first note that the infinite dihedral group is isomorphic to the free product of two cyclic groups of order 2, $\mathbb{Z}_{2} \ast \mathbb{Z}_{2}$, each of which is generated by two distinct parity transformations $\mathcal{P}_{0}$ and $\mathcal{P}_{\ell}$ that satisfy $\mathcal{P}_{0}^{2} = \mathcal{P}_{\ell}^{2} = 1$.
Defining the actions of these two parity transformations on $\mathbb{R}$ as $\mathcal{P}_{0}: y \mapsto -y$ and $\mathcal{P}_{\ell}: y \mapsto 2\ell - y$, whose fixed points are $y=0$ and $y=\ell$, respectively, we find that $\mathcal{P}_{0}$ and $\mathcal{P}_{0}\mathcal{P}_{\ell}$ correspond to the generators $\mathcal{R}$ and $\mathcal{T}$.
Noting that every element of $\mathbb{Z}_{2} \ast \mathbb{Z}_{2}$ is given by an alternating product of $\mathcal{P}_{0}$ and $\mathcal{P}_{\ell}$, we immediately see that the map $\mathbb{Z}_{2} \ast \mathbb{Z}_{2} \to D_{\infty}$ given by $(\mathcal{P}_{0}\mathcal{P}_{\ell})^{n} \mapsto \mathcal{T}^{n}$ and $(\mathcal{P}_{0}\mathcal{P}_{\ell})^{n}\mathcal{P}_{0} \mapsto \mathcal{T}^{n}\mathcal{R}$ is an isomorphism.
The second point to note is that $N$-dimensional unitary representation of $\mathbb{Z}_{2}$ is just given by an $N \times N$ hermitian unitary matrix \cite{Ohya:2011qu}.
Therefore, assigning two distinct hermitian unitary matrices to $\mathcal{P}_{0}$ and $\mathcal{P}_{\ell}$, we obtain the following $N$-dimensional unitary representations of $D_{\infty}$:
\begin{subequations}
\begin{align}
\mathbb{W}(\mathcal{T}^{n})
&= (\mathbb{U}_{0}\mathbb{U}_{\ell})^{n}, \label{eq:3.10a}\\
\mathbb{W}(\mathcal{T}^{n}\mathcal{R})
&= (\mathbb{U}_{0}\mathbb{U}_{\ell})^{n}\mathbb{U}_{0}, \label{eq:3.10b}
\end{align}
\end{subequations}
where $\mathbb{U}_{0}, \mathbb{U}_{\ell} \in U(N)$ are hermitian unitary matrices that satisfy $\mathbb{U}_{0}^{2} = \mathbb{U}_{\ell}^{2} = \mathbb{I}_{N}$.
These weight factors are nothing but those obtained in the case of free particle.

Now it is straightforward to derive the boundary conditions for the total Feynman kernel $\mathbb{K}$.
If the scalar kernel $K_{\mathbb{R}}$ is continuous and smooth at $x = 0$ and $\ell$, which is the case of $K_{\text{free}}$, we have
\begin{subequations}
\begin{align}
\mathbb{K}(0, y; T)
&= 	[\mathbb{I}_{N} + \mathbb{W}(\mathcal{R})]
	\sum_{n\in\mathbb{Z}}\mathbb{W}(\mathcal{T}^{n})K_{\mathbb{R}}(0, \mathcal{T}^{n}y; T), \label{eq:3.11a}\\
(\partial_{x}\mathbb{K})(0, y; T)
&= 	[\mathbb{I}_{N} - \mathbb{W}(\mathcal{R})]
	\sum_{n\in\mathbb{Z}}\mathbb{W}(\mathcal{T}^{n})(\partial_{x}K_{\mathbb{R}})(0, \mathcal{T}^{n}y; T), \label{eq:3.11b}\\
\mathbb{K}(\ell, y; T)
&= 	[\mathbb{I}_{N} + \mathbb{W}(\mathcal{RT})]
	\sum_{n\in\mathbb{Z}}\mathbb{W}(\mathcal{T}^{n})K_{\mathbb{R}}(\ell, \mathcal{T}^{n}y; T), \label{eq:3.11c}\\
(\partial_{x}\mathbb{K})(\ell, y; T)
&= 	[\mathbb{I}_{N} - \mathbb{W}(\mathcal{RT})]
	\sum_{n\in\mathbb{Z}}\mathbb{W}(\mathcal{T}^{n})(\partial_{x}K_{\mathbb{R}})(\ell, \mathcal{T}^{n}y; T), \label{eq:3.11d}
\end{align}
\end{subequations}
which follow from $\mathbb{W}(\mathcal{R}\mathcal{T}^{n}) = \mathbb{W}(\mathcal{R})\mathbb{W}(\mathcal{T}^{n})$, $\mathbb{W}(\mathcal{R}\mathcal{T}^{n+1}) = \mathbb{W}(\mathcal{R}\mathcal{T})\mathbb{W}(\mathcal{T}^{n})$ and $D_{\infty}$ symmetry \eqref{eq:3.3d}.
Making use of the unitary representations $\mathbb{W}(\mathcal{R}) = \mathbb{U}_{0}$ and $\mathbb{W}(\mathcal{RT}) = \mathbb{U}_{\ell}$, we immediately reproduce the scale-invariant boundary conditions $(\mathbb{I}_{N} - \mathbb{U}_{0})\mathbb{K}(0, y; T) = (\mathbb{I}_{N} + \mathbb{U}_{0})(\partial_{x}\mathbb{K})(0, y; T) = 0$ and $(\mathbb{I}_{N} - \mathbb{U}_{\ell})\mathbb{K}(\ell, y; T) = (\mathbb{I}_{N} + \mathbb{U}_{\ell})(\partial_{x}\mathbb{K})(\ell, y; T) = 0$, where we have used the orthogonality relations of hermitian unitary matrices $(\mathbb{I}_{N} \pm \mathbb{U}_{0})(\mathbb{I}_{N} \mp \mathbb{U}_{0}) = (\mathbb{I}_{N} \pm \mathbb{U}_{\ell})(\mathbb{I}_{N} \mp \mathbb{U}_{\ell}) = 0$.

Path integral representation of the total Feynman kernel \eqref{eq:3.2} is now easy.
Since path integral representation of $K_{\mathbb{R}}$ is a standard problem, we just assume that $K_{\mathbb{R}}$ is given by the standard configuration space path integral.
Then, if the bulk interactions are segment-independent, the path integral for systems on junctions of $N$ segments of equal length $\ell$ is given by the following $N \times N$ matrix-valued functional integral:
\begin{align}
\mathbb{K}(x, y; T)
&= 	\sum_{\mathcal{G} \in D_{\infty}}\mathbb{W}(\mathcal{G})
	\int_{x(0) = \mathcal{G}y}^{x(T) = x}\!\!\!\mathcal{D}x(t)
	\exp\left(i\int_{0}^{T}\!\!\!dt\,L\bigl(x(t), \Dot{x}(t)\bigr)\right), \label{eq:3.12}
\end{align}
where $\mathbb{W}(\mathcal{G})$ ($\mathcal{G} \in D_{\infty}$) is given in \eqref{eq:3.10a} \eqref{eq:3.10b}, $L\bigl(x(t), \Dot{x}(t)\bigr)$ is a generic $D_{\infty}$-invariant one-particle Lagrangian, and $\mathcal{D}x(t)$ is a $D_{\infty}$-invariant path integral measure on $\mathbb{R}$.

\section{Conclusions} \label{sec:4}
In this paper we have studied path integral description of quantum mechanics on junctions of $N$ segments of equal length $\ell$.
We have seen that scale-invariant subfamily of boundary conditions is well translated into the problem of weight factors in path integral.
Provided the bulk Hamiltonian is segment-independent, the weight factors are generally given by $N$-dimensional unitary representations of the infinite dihedral group $D_{\infty} \cong \mathbb{Z}_{2} \ast \mathbb{Z}_{2}$; that is, in this case the problem of boundary conditions boils down to the problem of unitary representations.
These two problems are in one-to-one correspondence with each other, since both two are solved once two hermitian unitary matrices $\mathbb{U}_{0}$ and $\mathbb{U}_{\ell}$ are specified.

It should be noted that our result \eqref{eq:3.12} does not cover the whole self-adjoint domain of Hamiltonian operator.
For non-scale-invariant case, the boundary S-matrices generally depend on the particle momentum such that the total Feynman kernel cannot be factorized as $(\text{scalar kernel}) \times (\text{$N \times N$ constant matrix})$ in the configuration space path integral.
As discussed in the previous work \cite{Ohya:2011qu}, the key to understanding the non-scale-invariant boundary conditions would be to generalize the weight factors in the phase space path integral.
We hope to address this issue elsewhere.

\appendix
\makeatletter
\def\@seccntformat#1{Appendix\ \csname the#1\endcsname\quad}
\makeatother
\section{Feynman kernel for a free particle} \label{appendix:A}
In this appendix we first solve the Schr\"odinger equation with scale-invariant boundary conditions \eqref{eq:2.2a} \eqref{eq:2.2b} and then derive the Feynman kernel \eqref{eq:2.3} by using the complete orthonormal basis of energy eigenfunctions and the argument principle.

\begin{figure}[t]
\centerline{
\begin{minipage}[t]{.48\textwidth}
\begin{center}
\includegraphics{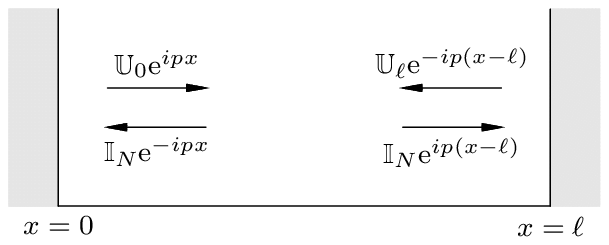}
\caption{One-particle boundary S-matrices $\mathbb{U}_{0}$ and $\mathbb{U}_{\ell}$.}
\label{fig:4}
\end{center}
\end{minipage}
\hspace{.05\textwidth}
\begin{minipage}[t]{.48\textwidth}
\begin{center}
\includegraphics{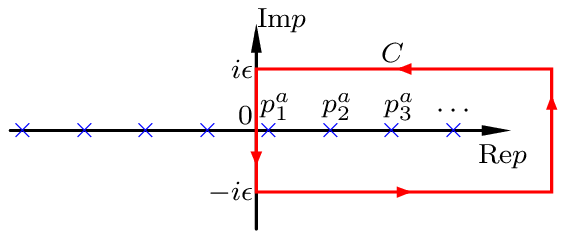}
\caption{Simple poles ($\color{blue}\times$) and integration contour $C$.}
\label{fig:5}
\end{center}
\end{minipage}
}
\end{figure}

The general solution to the Schr\"odinger equation $\mathbb{H}_{\text{free}}\Vec{\psi}(x) = E\Vec{\psi}(x)$ for positive energy $E > 0$ is given by a linear combination of plain waves
\begin{align}
\Vec{\psi}(x; p)
&= 	\Vec{A}_{0}\mathrm{e}^{-ipx} + \Vec{A}_{\ell}\mathrm{e}^{-ip(\ell - x)}, \quad
	p = \sqrt{E} > 0, \label{eq:A.1}
\end{align}
where $\Vec{A}_{0}$ and $\Vec{A}_{\ell}$ are $N$-component coefficient column vectors.
Substituting the general solution to the boundary conditions \eqref{eq:2.2a} \eqref{eq:2.2b} we get the following eigenvalue equation:
\begin{align}
\mathbb{M}(p)
\begin{pmatrix}
\Vec{A}_{0} \\
\Vec{A}_{\ell}
\end{pmatrix}
&= 	\begin{pmatrix}
	\Vec{A}_{0} \\
	\Vec{A}_{\ell}
	\end{pmatrix}, \label{eq:A.2}
\end{align}
where $\mathbb{M}(p)$ is a $2N \times 2N$ unitary matrix given by
\begin{align}
\mathbb{M}(p)
= 	\begin{pmatrix}
	0 		& \mathbb{U}_{\ell} \\
	\mathbb{U}_{0} 	& 0
	\end{pmatrix}
	\mathrm{e}^{ip\ell}. \label{eq:A.3}
\end{align}
It is worth mentioning here the physical meaning of the hermitian unitary matrices $\mathbb{U}_{0}$ and $\mathbb{U}_{\ell}$.
By using the relations $\Vec{A}_{\ell} = \mathbb{U}_{0}\mathrm{e}^{ip\ell}\Vec{A}_{0}$ and $\Vec{A}_{0} = \mathbb{U}_{\ell}\mathrm{e}^{ip\ell}\Vec{A}_{\ell}$, which follow from \eqref{eq:A.2}, the solution \eqref{eq:A.1} can be rewritten as $\Vec{\psi}(x; p) = (\mathbb{I}_{N}\mathrm{e}^{-ipx} + \mathbb{U}_{0}\mathrm{e}^{ipx})\Vec{A}_{0} = (\mathbb{I}_{N}\mathrm{e}^{ip(x - \ell)} + \mathbb{U}_{\ell}\mathrm{e}^{-ip(x - \ell)})\Vec{A}_{\ell}$.
These expressions imply that $\mathbb{U}_{0}$ and $\mathbb{U}_{\ell}$ play the roles of boundary S-matrices at $x = 0$ and $\ell$, whose $ij$-component gives the transmission amplitude (or reflection amplitude if $i = j$) for a particle traveling from the $j$th segment to the $i$th segment; see figure \ref{fig:4}.

Let us next discuss the complete orthonormal set of energy eigenfunctions.
The condition \eqref{eq:A.2} implies that the eigenvectors and eigenvalues of $\mathbb{M}(p)$ will respectively determine the coefficient vectors $\Vec{A}_{0}$, $\Vec{A}_{\ell}$ and the energy spectrum via quantization condition $(\text{eigenvalue of $\mathbb{M}(p)$}) = 1$.
Let $\mathrm{e}^{i\theta^{a}}$ ($0 \leq \theta^{a} < 2\pi$) and $\Vec{A}^{a} = \bigl(\begin{smallmatrix} \Vec{A}_{0}^{a} \\ \Vec{A}_{\ell}^{a}\end{smallmatrix}\bigr)$ ($a = 1, \cdots, 2N$) be the $a$th eigenvalue and its corresponding normalized eigenvector of the unitary matrix $\bigl(\begin{smallmatrix} 0 & \mathbb{U}_{\ell} \\ \mathbb{U}_{0} & 0\end{smallmatrix}\bigr)$ that satisfy the eigenvalue equation $\bigl(\begin{smallmatrix} 0 & \mathbb{U}_{\ell} \\ \mathbb{U}_{0} & 0\end{smallmatrix}\bigr)\Vec{A}^{a} = \mathrm{e}^{i\theta^{a}}\Vec{A}^{a}$, orthonormality $\Vec{A}^{a\dagger}\Vec{A}^{b} = \delta^{ab}$ and completeness $\sum_{a=1}^{2N}\Vec{A}^{a}\Vec{A}^{a\dagger} = \mathbb{I}_{2N}$.
The normalized energy eigenfunction is then given by $\Vec{\psi}^{a}(x; p^{a}_{n}) = \frac{1}{\sqrt{\ell}}(\Vec{A}_{0}^{a}\mathrm{e}^{-ip^{a}_{n}x} + \Vec{A}_{\ell}^{a}\mathrm{e}^{-ip^{a}_{n}(\ell - x)})$ with $p^{a}_{n} = \tfrac{2n\pi - \theta^{a}}{\ell} > 0$ ($n= 1,2,\cdots$) being the positive roots of the quantization condition $\mathrm{e}^{ip\ell + i\theta^{a}} = 1$.\footnote{For the sake of simplicity we here assume that $\theta^{a} \neq 0$ for any $a$, which guarantees the absence of constant zero-modes in the spectrum.
For the case of the presence of zero-modes, one has to modify the integration contour $C$ in figure \ref{fig:5} by changing the leftmost vertical line to an infinitesimal semicircle.
However, the result \eqref{eq:A.7} remains unchanged regardless of the presence or absence of zero-modes.}
These normalized energy eigenfunctions provide the complete orthonormal basis of the system such that the Feynman kernel is expanded as $\mathbb{K}(x, y; T) = \sum_{a=1}^{2N}\sum_{n=1}^{\infty}\Vec{\psi}^{a}(x; p^{a}_{n})\mathrm{e}^{-iT(p^{a}_{n})^{2}}\Vec{\psi}^{a\dagger}(y; p^{a}_{n})$, which can be written as follows:
\begin{align}
\mathbb{K}(x, y; T)
&= 	\frac{1}{\ell}\sum_{a=1}^{2N}\sum_{n=1}^{\infty}
	\begin{pmatrix}
	\mathrm{e}^{ip^{a}_{n}x}\mathbb{I}_{N} \\
	\mathrm{e}^{ip^{a}_{n}(\ell - x)}\mathbb{I}_{N}
	\end{pmatrix}^{\dagger}
	\mathbb{P}^{a}
	\begin{pmatrix}
	\mathrm{e}^{ip^{a}_{n}y}\mathbb{I}_{N} \\
	\mathrm{e}^{ip^{a}_{n}(\ell - y)}\mathbb{I}_{N}
	\end{pmatrix}
	\mathrm{e}^{-iT(p^{a}_{n})^{2}}, \label{eq:A.4}
\end{align}
where $\mathbb{P}^{a} = \Vec{A}^{a}\Vec{A}^{a\dagger}$ is a hermitian projector onto the $a$th eigenspace for the unitary matrix $\mathbb{M}(p)$.
(Note that the energy eigenfunction can be written as $\Vec{\psi}^{a}(x; p) = \frac{1}{\sqrt{\ell}}\bigl(\begin{smallmatrix} \mathrm{e}^{ipx}\mathbb{I}_{N} \\ \mathrm{e}^{ip(\ell - x)}\mathbb{I}_{N}\end{smallmatrix}\bigr)^{\dagger}\Vec{A}^{a}$.)

Now we wish to switch from the summation over the energy spectrum to the summation over (classical) trajectories, which can be achieved via the following contour integral (i.e. argument principle):
\begin{align}
\sum_{n=1}^{\infty}f(p^{a}_{n})
&= 	\oint_{C}\frac{dp}{2\pi i}f(p)\frac{d}{dp}\log(1 - \mathrm{e}^{i(p\ell + \theta^{a})}) \nonumber\\
&= 	\ell
	\left[
	\sum_{m=1}^{\infty}\int_{0 + i\epsilon}^{\infty + i\epsilon}\!\frac{dp}{2\pi}\,
	f(p)\mathrm{e}^{im(p\ell + \theta^{a})}
	+ \sum_{m=0}^{\infty}\int_{0 - i\epsilon}^{\infty - i\epsilon}\!\frac{dp}{2\pi}\,
	f(p)\mathrm{e}^{-im(p\ell + \theta^{a})}
	\right] \nonumber\\
&\hspace{-1.4ex}\stackrel{\epsilon \to 0_{+}}{\to}
	\ell\sum_{m\in\mathbb{Z}}\int_{0}^{\infty}\!\frac{dp}{2\pi}\,
	\mathrm{e}^{im(p\ell + \theta^{a})}f(p),
	\quad (\text{$f$: test function}), \label{eq:A.5}
\end{align}
where the integration contour $C$ is chosen to enclose all the positive roots of the quantization condition in a counterclockwise direction; see figure \ref{fig:5}.
The first equality just follows from the residue theorem and the second equality the geometric series expansion:
\begin{align}
\frac{d}{dp}\log(1 - \mathrm{e}^{i(p\ell + \theta^{a})})
&= 	\begin{cases}
	\displaystyle
	-i\ell\sum_{m=1}^{\infty}\mathrm{e}^{im(p\ell + \theta^{a})} 	& \text{for $\mathrm{Im}\,p > 0$}, \\
	\displaystyle
	+i\ell\sum_{m=0}^{\infty}\mathrm{e}^{-im(p\ell + \theta^{a})} 	& \text{for $\mathrm{Im}\,p < 0$}.
	\end{cases} \label{eq:A.6}
\end{align}
Since there are no longer poles in the integrand, we can take the limit $\epsilon \to 0_{+}$ in the third line of \eqref{eq:A.5}.

Now, with the aid of the formula \eqref{eq:A.5} the Feynman kernel \eqref{eq:A.4} is cast into the following form:
\begin{align}
\mathbb{K}(x, y; T)
&= 	\sum_{m\in\mathbb{Z}}
	\int_{0}^{\infty}\!\frac{dp}{2\pi}
	\begin{pmatrix}
	\mathrm{e}^{ipx}\mathbb{I}_{N} \\
	\mathrm{e}^{ip(\ell - x)}\mathbb{I}_{N}
	\end{pmatrix}^{\dagger}
	\mathbb{M}^{m}(p)
	\begin{pmatrix}
	\mathrm{e}^{ipy}\mathbb{I}_{N} \\
	\mathrm{e}^{ip(\ell - y)}\mathbb{I}_{N}
	\end{pmatrix}
	\mathrm{e}^{-iTp^{2}}, \label{eq:A.7}
\end{align}
which follows from the spectral decomposition $\mathbb{M}^{m}(p) = \sum_{a=1}^{2N}\mathrm{e}^{im(p\ell + \theta^{a})}\mathbb{P}^{a}$.
By using the definition \eqref{eq:A.3}, the $m$th power of unitary matrix $\mathbb{M}(p)$ is easily calculated with the result
\begin{align}
\mathbb{M}^{m}(p)
&= 	\begin{cases}
	\displaystyle
	\begin{pmatrix}
	(\mathbb{U}_{\ell}\mathbb{U}_{0})^{n} 	& 0 \\
	0 								& (\mathbb{U}_{0}\mathbb{U}_{\ell})^{n}
	\end{pmatrix}
	\mathrm{e}^{ip2n\ell}
	& \text{for $m = 2n$}, \\[1.5em]
	\displaystyle
	\begin{pmatrix}
	0 											& (\mathbb{U}_{\ell}\mathbb{U}_{0})^{n}\mathbb{U}_{\ell} \\
	(\mathbb{U}_{0}\mathbb{U}_{\ell})^{n}\mathbb{U}_{0} 	& 0
	\end{pmatrix}
	\mathrm{e}^{ip(2n+1)\ell}
	& \text{for $m = 2n+1$}, \label{eq:A.8}
	\end{cases}
\end{align}
where $n$ is an integer.
Substituting \eqref{eq:A.8} into \eqref{eq:A.7} one immediately sees that the total Feynman kernel is recast into the following form:
\begin{align}
\mathbb{K}(x, y; T)
&= 	\sum_{n\in\mathbb{Z}}(\mathbb{U}_{0}\mathbb{U}_{\ell})^{n}
	\int_{-\infty}^{\infty}\!\frac{dp}{2\pi}
	\exp
	\left(
	iT\left[p\left(\frac{x - y + 2n\ell}{T}\right) - p^{2}\right]
	\right) \nonumber\\
& 	+
	\sum_{n\in\mathbb{Z}}(\mathbb{U}_{0}\mathbb{U}_{\ell})^{n}\mathbb{U}_{0}
	\int_{-\infty}^{\infty}\!\frac{dp}{2\pi}
	\exp
	\left(
	iT\left[p\left(\frac{x + y + 2n\ell}{T}\right) - p^{2}\right]
	\right). \label{eq:A.9}
\end{align}
By performing the momentum integration, we obtain the result \eqref{eq:2.3}.

\bibliographystyle{utphys}
\bibliography{Bibliography}

\end{document}